# Feasibility of Characterizing Subsurface Brines on Ceres by Electromagnetic Sounding

Robert Grimm[1], Julie Castillo-Rogez[2], Carol Raymond[2], Andrew R. Poppe[3]

[1]Planetary Science Directorate, Southwest Research Institute, Boulder, Colorado, USA

[2]Jet Propulsion Laboratory, California Institute of Technology, Pasadena, California, USA

[3]Space Sciences Laboratory, University of California, Berkeley, California, USA

Corresponding author: Robert Grimm, grimm@boulder.swri.edu

## Abstract

Electromagnetic sounding of Ceres is straightforward using the solar wind as a source. The depths to a deep global brine or mud layer and shallow briny intrusions can be assessed simultaneously.

## 1. Introduction

Dwarf planet Ceres is the largest object in the main belt and the most water-rich object in the inner solar system (in relative abundance of water to rock). Ceres had sufficient water and silicates (including radioisotopes) to host a deep ocean in its past, leading to a layered interior structure with a high degree of aqueous alteration (Ammannito et al., 2016, Ermakov et al. 2017). The Dawn mission revealed evidence of recent and possibly ongoing geologic activity on Ceres (De Sanctis et al. 2020) and the presence of liquid was inferred below an ice-rich crust (Fu et al. 2017; Scully et al. 2020, Raymond et al. 2020). Recent brine-driven exposure of material onto Ceres' surface can be found at Occator crater (Nathues et al. 2020) and Ahuna Mons (Ruesch et al. 2016). These multiple lines of evidence for deep liquid and long-lived heat sources call for categorization of Ceres as an ocean world (Castillo-Rogez, 2020).

Determination of the depth to liquid water within Ceres—specifically below Occator crater—is one objective of a recent Planetary Mission Concept Study (Castillo-Rogez and Brophy, 2020). This sample-return mission includes a landed phase of several days, allowing the opportunity for surface geophysics to provide tighter constraints on interior water over those derived from orbit. Passive seismology could be useful, but there is insufficient information at present on the strength, distribution, and recurrence of volcanotectonic or impact seismic sources suitable for analysis by a single, short-duration station. Without large resources, active-source seismology, radar, or low-frequency electromagnetics cannot penetrate the required tens of kilometers of icy crust. However, passive low-frequency electromagnetic (EM) methods are not only tractable, but optimal for this task. First, the solar-wind EM source for sounding at Ceres is already well-understood, including measurement of both the power spectral density in the inner solar system and the overall amplitude dependence with heliocentric distance. Second, water with even a small quantity of dissolved salts is a strong electrical conductor compared to ice or rock: finding conductors is the essential ability of inductive EM (e.g., the familiar metal detector). The value of inductive EM for characterizing subsurface water on Mars and Europa has been studied by Grimm (2002) and Grimm et al. (2020), respectively.

In this note, we apply the same approaches to Ceres. We describe how EM sounding can be performed and derive an error budget for a representative experiment. We specify a likely range of internal structures and evaluate their EM responses. We conclude that the distances to both a deep brine-rich reservoir and a shallow brine-filled body feeding eruptions can be determined simultaneously, and furthermore that such sounding can be implemented with high confidence and low resources.

## 2. Electromagnetic Sounding

A changing magnetic field induces eddy currents to flow in any body with finite electrical conductivity. These currents have secondary magnetic fields that oppose the primary, such that the net magnetic field decays with 1/e folding ("skin") depth $\sqrt{2}/\mu\omega\sigma$, where $\mu$ is the magnetic permeability, $\omega$ is the angular frequency, and $\sigma$ is the electrical conductivity. The skin-depth effect is the essence of EM sounding that allows frequency-dependent measurements of EM fields to be translated into conductivity as a function of depth.

Although it is possible to fit EM fields directly, solid-earth geophysics commonly translates these data to proxies that have physical meaning. Here we use the apparent conductivity $\sigma_a$, which is the conductivity of a half-space having the same response as the target. Then EM sounding can be considered in terms of using the skin-depth effect to invert $\sigma(z)$ from $\sigma_a(f)$, where $z$ is depth and $f$ is frequency. Alternatively, the (complex) impedance $Z$ ($\sigma_a = \mu\omega/|Z|^2$) can be the most descriptive when considering measurement methods: the familiar Ohm's Law $Z = V/I$ calls for two independent quantities to determine the impedance, and the same applies to EM sounding.

There are two approaches to determining $\sigma_a$ or $Z$ that are relevant to Ceres (see Grimm and Delory, 2012; Banerdt et al., 2014). The Transfer Function (TF) compares the magnetic field at a planetary surface (the sum of source and induced fields) to the distant source field. This transfer function was implemented between the distantly orbiting Explorer 35 satellite and the Apollo 12 lunar surface magnetometer (e.g., Sonett, 1982). The Galileo induction studies can be considered a special case of the TF, wherein very low-frequency source fields are known and modeled a priori (e.g., Khurana, 1997). It is important to recognize that if the source field is not known independently, one magnetometer is insufficient to perform EM sounding.

In the Magnetotelluric Method (MT), the ratio of the electric field *E* and the magnetic field *B* is proportional to the impedance, where *E* and *B* are in fact measured orthogonally and parallel to the surface. Because *E* supplies the required second piece of information, the source field does not need to be known. Tensor impedance measurements (using both horizontal components of both *E* and *B*) further enable anisotropy, or directionality of targets, to be determined. MT is among the most highly developed and widely applied EM method for investigations of Earth's crust and upper mantle (e.g., Simpson and Bahr, 2005; Berdichevsky and Dmitriev, 2010; Chave and Jones, 2012).

## 3. Measurement

We analyze the performance of a magnetotelluric sounder on Ceres by the same approach used by Grimm et al. (2020) for a similar instrument on Europa. The closely related Lunar Magnetotelluric Sounder has been selected for flight to the Moon under NASA's Commercial Lunar Payload Services program. Key characteristics of an instrument for Ceres are shown in **Fig.**

**1**. The minimum frequency considered is $6.1 \times 10^{-5}$ Hz, corresponding to one-half of Ceres 9.1-hr rotation period. This assures that all measurements are made in daylight, because the electrodes rely on coupling to the environment via photoelectrons (Bonnell et al., 2008). The maximum frequency 128 Hz is the Nyquist frequency of the MAVEN magnetometer used for the Europa prototype (Connerney et al., 2015).

The magnetometer and electrometer noise floors are derived from MAVEN (Connerney et al., 2015) and THEMIS (Bonnell et al., 2008) respectively. The magnetometer performance is about four times worse than MAVEN because a smaller fluxgate sensor is used. The electrometer performance is about ten times better than THEMIS because the electrodes are assumed to be separated by 200 m. A spring launcher derived from the Europa design can readily deploy an electrode and connecting cable to 100 m in the low gravity of Ceres; the electric field is measured from the voltage difference between oppositely deployed electrodes. We assume that signal integration can take place over 5 Ceres days (about 50 hr), based on the planned duration of the landed mission (Castillo-Rogez and Brophy, 2020). This has the effect of pushing the noise floor down in proportion to the square root of the number of cycles measured, although we require 30 cycles to achieve unit integration gain. The raw noise floors and the integration gain are shown separately in Fig. 1.

The source magnetic field is taken to be that of the solar wind, scaled to Ceres. We first determined the average spectrum near the Earth by averaging five years of data from the ARTEMIS spacecraft (Angelopoulos, 2014) when in the solar wind. For modeling purposes, we fit this log-log curve to three line segments (this leads to some slope breaks in Fig. 1). Conservatively, we reduce all amplitudes by a factor of five, which is approximately the reduction seen at −1 standard deviation in the cislunar environment. Next, we scale the solar-wind amplitude to Ceres as $A = \sqrt{1 + R^2}/R^2$ (Klein et al., 1987), where A is the normalized amplitude and R is the normalized heliocentric distance. For R = 2.76 AU, A = 0.385. The adopted Ceres magnetic-field spectrum (Fig. 1) is therefore the spectrum measured near Earth scaled by a factor of 0.077.

The E-field is calculated from the input B-field and the reference electrical conductivity (impedance) of Ceres as described below. This is the minimum electric field and defines the measurement requirement. Plasma waves may cause other electric fields, but they can be separated using the frequency-dependent amplitude and phase relations that planetary induction must satisfy. This formulation also does not account for solar-wind confinement of the induced field on the dayside (e.g., Sonett, 1982), so in fact the input B-fields are minima. Faraday's Law holds nonetheless, so the computed E varies proportionally and the MT ratio still applies.

Since $\sigma_a \propto (B/E)^2$, the variance in the apparent conductivity $\varsigma^2_{\sigma a}$ follows from the variances in the electric and magnetic fields, $\varsigma^2_E$ and $\varsigma^2_B$, respectively, as

$$\frac{\varsigma^2_{\sigma_a}}{\sigma_a^2} = 4\left(\frac{\varsigma^2_E}{E^2} + \frac{\varsigma^2_B}{B^2}\right) \qquad (1)$$

The joint signal-to-noise ratio SNR = N($\sigma_a/\varsigma_{\sigma a}$), where N is the integration gain (Fig. 1). We find that SNR is >1 for all frequencies <60 Hz, with a maximum of 80 at 20 mHz. The median SNR is 24. The errors $\varsigma_{\sigma a}$ are plotted on the reference model in Figs. 2 and 3.

By comparison, the SNR for the Apollo 12 – Explorer 35 transfer lunar transfer function was ~20 over $10^{-5}$ to $10^{-3}$ Hz (Hobbs et al., 1983). However, at higher frequencies—where wavelengths in the plasma are comparable to the planetary circumference—distortions are introduced into the transfer function. While there has been some success in treating these effects up to 30 mHz (Sonett et al., 1972), the best inversions for the conductivity structure of the Moon restricted the frequency to ≤1 mHz (Hood et al., 1982; Khan et al., 2006). Assuming that the useful bandwidth scales inversely as planetary radius, plasma distortion to the transfer function at Ceres will appear >4 mHz and will be intractable >100 mHz. This will not affect the ability of TF to probe for a deep brine layer, although shallow brine conduits will be undetectable (see below).

## 4. Modeling

We specify a series of 1D (layered) models $\sigma(z)$ for Ceres and calculate the frequency-dependent apparent conductivities $\sigma_a(f)$ using the method described by Grimm and Delory (2012). This uses a plane-layer propagator (Wait, 1970) with spherical transformation (Wiedelt, 1972) and allows an arbitrarily discretized radial conductivity structure. This model effectively applies to the dipole response of a sphere in a vacuum and does not treat dayside solar-wind confinement of the induced field. This is unnecessary for two reasons. First, MT is insensitive to the upstream field properties (see above). Second, the confinement factor can be computed from the vacuum response (see Sonett, 1982, Eqn. 33 and Fig. 11), such that all the magnetic fields for TF can be derived. One-standard-deviation error bars were derived from the joint errors in the electric and magnetic field measurements (MT method) described above.

The reference structure has a 35 km thick crust with conductivity $10^{-4}$ S/m that overlies a brine-rich layer with conductivity 1 S/m extending to 100 km depth. For convenience, we call the brine reservoir an "ocean," although it could contain substantial silicates. The crustal thickness is rounded from Ermakov et al. (2017). The crustal conductivity is the maximum DC subeutectic conductivity of ice formed from an NaCl-saturated solution (Grimm et al., 2008). It will be seen that this value, and hence all smaller values, have little effect on the results. The ocean conductivity is varied between 20 S/m (approximately a saturated chloride solution, CRC, 2008, p. 5-72) and 0.2 S/m (5 S/m seawater in 20% porosity, Archie's Law with cementation exponent 2), with the reference value near the logarithmic mean. Below that, a phyllosilicate mantle is assigned a conductivity of $10^{-2}$ S/m, representative of consolidated shale (Telford et al., 1990, p. 290). In order to assess discrimination of shallow vs deep conductors, a briny sill (0.2 km, 0.1 S/m) is inserted at 2 km depth. This is included in the reference model and is used for all parameter studies that follow. The single layer can be considered a proxy for multiple thinner and brinier intrusions (e.g., 10 sills in series with thickness 10 m and conductivity 20 S/m). The sill is chosen for convenience in representing shallow water bodies because it can be incorporated in a 1D model. A dike is more plausible (Raymond et al., 2020; Scully et al., 2020) but requires a 3D model: representative 3D calculations are given in the Supplementary Material.

The actual and apparent conductivities for the reference model, and for models successively deleting the sill and then the ocean, are shown in **Fig. 2**. First consider the dry structure (red curve). At the highest frequencies, $\sigma_a$ is equal to the true conductivity of the crust. These signals are entirely absorbed in the crust. Below 1 Hz, the response trends toward a −45° log-log slope. This is diagnostic of signals that are sensing a strong conductor and follows $\sigma_a = 1/\mu\omega h^2$, where $h$ is the depth to the conductor. For strong, thick, conductors, the depth to the conductor can be directly read off from any point on this "h-line" (Berdichevsky and Dmitriev, 2010). In this first case, however, the slope begins shallowing as lower frequencies penetrate farther into the phyllosilicate mantle and begin to asymptote towards its actual conductivity $10^{-2}$ S/m. Before this is fully realized, the signals have completely penetrated Ceres and $\sigma_a$ trends back to another h-line that represents a geometric limit: below ~ $3\times10^{-4}$ Hz, the EM response simply indicates the presence of a sphere (Ceres) in free space without any information on its composition (dashed line in Fig. 2).

When the ocean is present at the bottom of the crust (green curves), the high-frequency response to the uppermost layer is the same, but the h-line continues across four decades of frequency. Measurements anywhere along this extensive h-line are sufficient to determine the depth to brine. The EM signals do not begin penetrating into the underlying mantle until the lowest frequencies.

The full reference model with both sill and ocean (blue curves) introduces a new h-line at high frequency, whose higher apparent conductivity reveals the shallower depth of the sill. The transition between the signatures of the sill and the ocean is around 1 Hz. The segment 0.2 to 3 Hz has +45° log-log slope and is given by $\sigma_a = \mu\omega S^2$, where $S = \sigma H$ is the conductance of the overlying sill of thickness $H$. This "S-line" (Berdichevsky and Dmitriev, 2010) indicates that the conductivity and thickness of the sill cannot be determined separately. If the sill was much thicker, $\sigma_a$ would asymptote to the sill's actual $\sigma$ at lower frequency and the S-line could then be used to determine the sill's thickness. At lower frequencies here, the ocean's h-line signature is still evident.

In summary, the reference model broadly indicates that the ocean depth can be determined at frequencies between $6 \times 10^{-5}$ Hz and 1 Hz if no sill is present. Otherwise, the ocean's signature is still evident up to 0.3 Hz and the depth to the sill is determined above 3 Hz.

In **Fig. 3**, we explore the effects of six parameter variations (crust thickness, ocean thickness and conductivity, sill depth, thickness, and conductivity) from the reference model. There is little sensitivity to ocean thickness >30 km (depth 65 km) because signals at useful frequencies are absorbed by the saltwater. Note that here the calculation for zero ocean thickness still includes the overlying sill. Ocean conductivities lower than the reference value—due to lower salt content or a higher silicate fraction—can be distinguished, but the curves for near-eutectic compositions merge as signal penetration is diminished. H-lines for the depth to the top of the ocean (crust thickness) and the depth to the top of the sill are well-developed. Test values for ocean depth are closely spaced because there are already constraints (Ermakov et al., 2017), and the small expected measurement errors in this band suggest that the shell thickness can be determined to within a few km. The families of S-lines for sill thickness and conductivity illustrate the equivalence (ambiguity) in separating these parameters.

## 5. Concluding Discussion

The subsurface water reservoirs of Ceres may hold important clues to the evolution, differentiation, and habitability of ocean worlds. Electromagnetic sounding is the best geophysical approach for assessing not just a large, deep, brine reservoir, but also liquid in shallow intrusions that may be the sources of recent extrusion. An orbiter with high apoapsis and a lander can together perform basic sounding using the magnetic transfer function method. However, the transfer function is likely be limited to <0.1 Hz due to plasma effects. At these frequencies, the ocean can be detected but not shallow intrusions. This method requires concurrent observations by two flight elements, which increases mission cost. In contrast, the magnetotelluric method measures both electric and magnetic fields at the lander and does not require an orbiter. It is largely insensitive to plasma effects and can distinguish different shallow intrusive structures by their anisotropic responses. A magnetotelluric instrument is baselined in the payload of a Ceres lander and sample-return mission study under consideration for the 2023-2032 Planetary Science and Astrobiology Decadal Survey.

## Acknowledgements

This work was funded by NASA grants 80NSSC17K0769 (AP) and 80NSSC19K0609 (RG). We thank Mike Sori, Simone Marchi, and David Stillman for helpful discussion and review. RG is grateful to the Europa / Lunar Magnetotelluric Sounder Team for their contributions to realizing this technique in planetary exploration.

Figures.

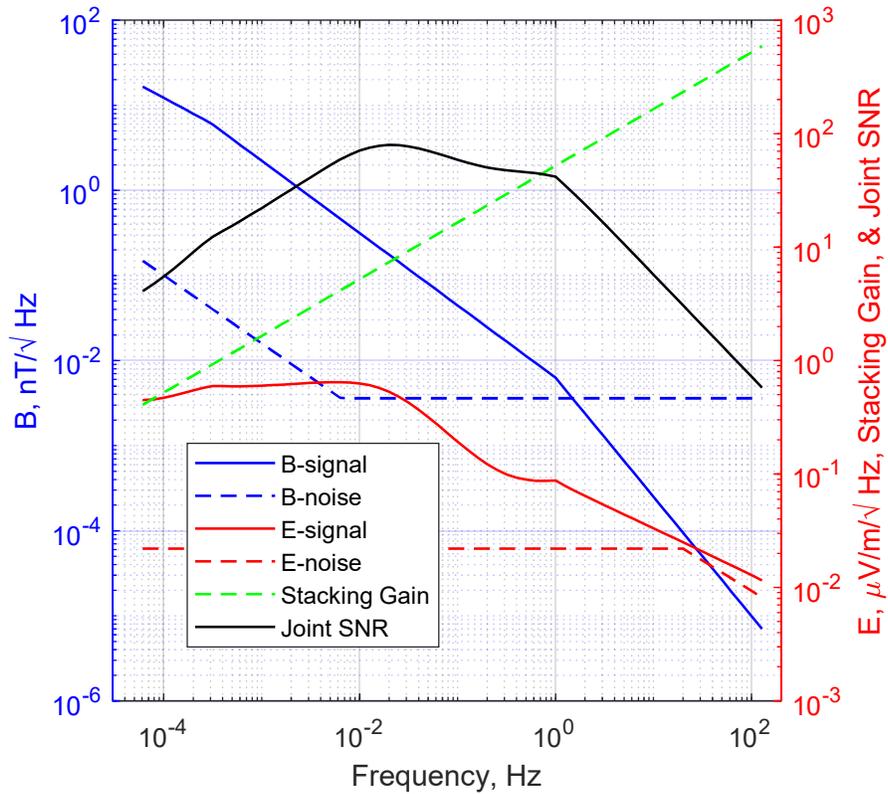

**Fig. 1**. Analysis of signals and noise for magnetotelluric sounding of Ceres (see text). Solar wind magnetic field B is reduced by 1 standard deviation from the expected mean value. Electric field E is calculated from the reference interior structure. Noise floors for magnetometer and electrometer are modified from MAVEN and THEMIS, respectively. Stacking gain assumes integration for 5 Ceres days (daylight only). The joint SNR is derived from the product of the stacking gain and the sum of the variances of the two sensors, and has a median value of 24.

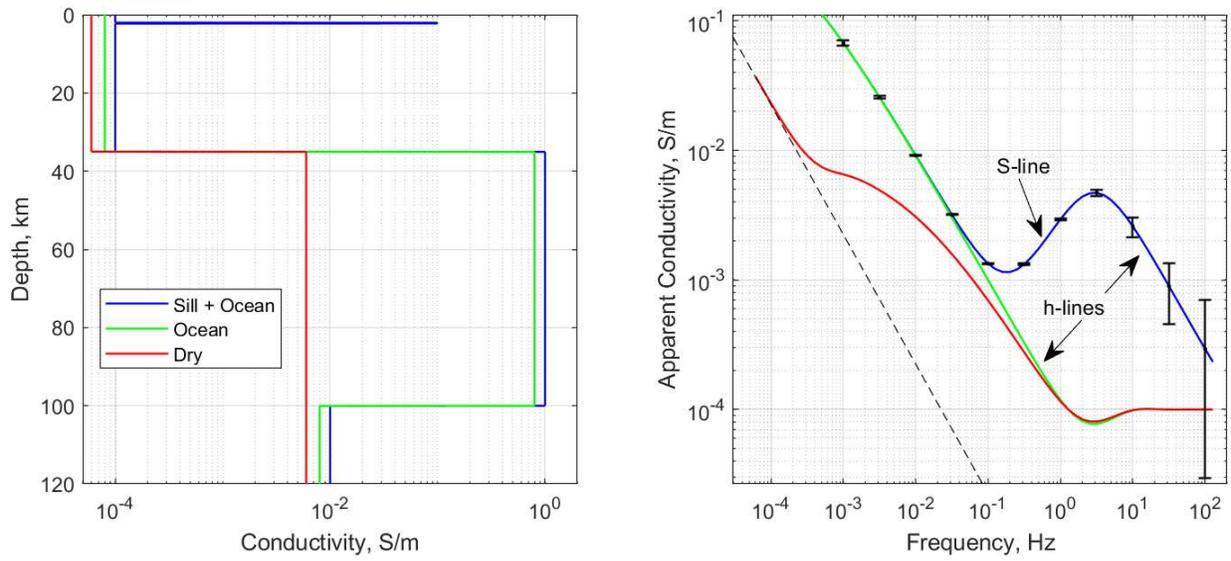

**Fig. 2**: Basic interior models (left, displaced for clarity) and EM responses (right) for Ceres. See text for discussion. Error bars are derived from Fig. 1.

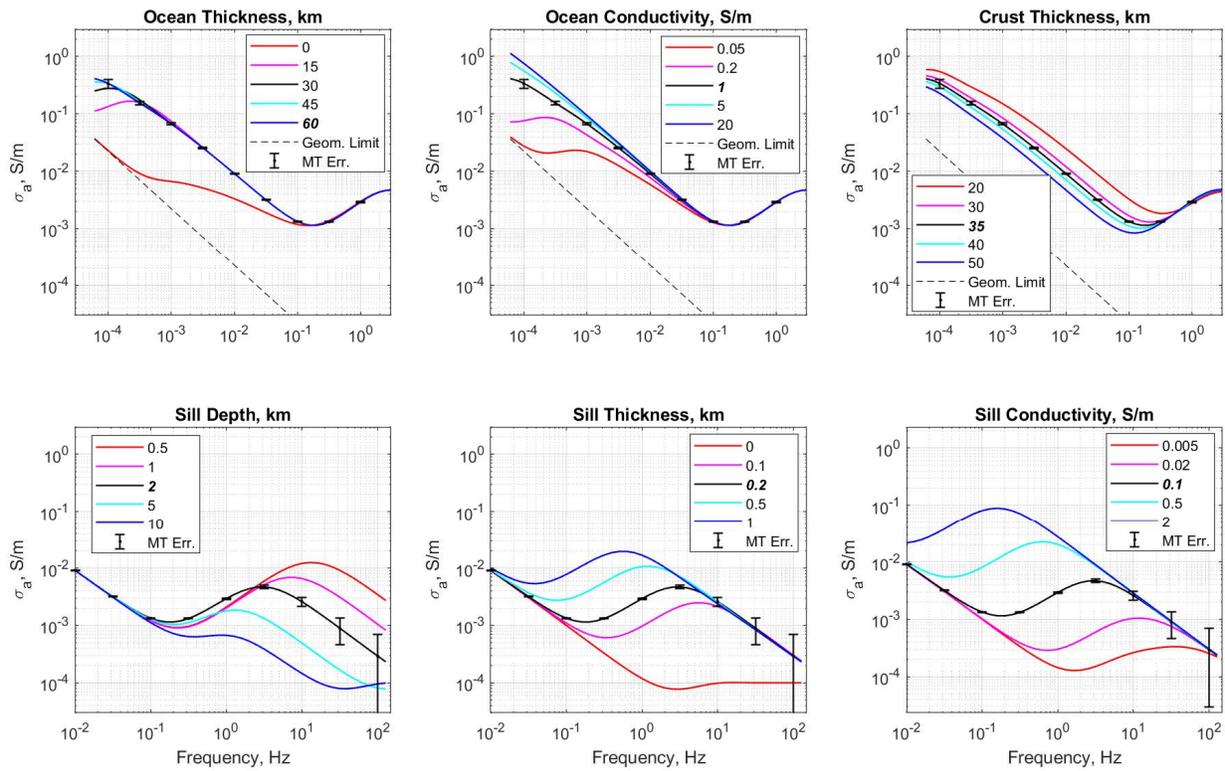

**Fig. 3**: Parametric study of EM sounding of Ceres. Note different abscissa ranges in top and bottom rows. Crustal thickness is well-characterized <0.1 Hz, but ocean thickness and conductivity are ambiguous. Similarly, the depth to a shallow conductor (sill) is readily measured >1-10 Hz, but only its thickness-conductivity product can be determined.

Supplementary Material For

**Feasibility of Characterizing Subsurface Brines on Ceres**

**by Electromagnetic Sounding**

by R.E. Grimm et al.

We evaluated a small set of 3D induction models using the finite-element method, similar to our effort for Europa (Grimm et al., 2020). The Frequency Domain formulation in the Comsol™ RF module was used. The MT impedance is a 2x2 matrix, giving orthogonal induced E-fields as functions of both x- and y- components of the source magnetic field. In practice, the impedance tensor is rotated to eliminate the diagonal components so that $\mu E_x = Z_{xy} B_y$ and $\mu E_y = Z_{yx} B_x$, as a function of frequency. These cases can be evaluated separately numerically; if the lander and the target lie on the y-axis, then $Z_{yx}$ is the Transverse Magnetic (TM) mode and $Z_{xy}$ is the Transverse Electric (TE) mode (**Fig. S1**).

The model domain is a rectangular prism 20 x 20 x 35 km. This captures the reference crustal thickness vertically and is much larger than the 2-km distance between the target and the observation point, so that field perturbations are small where the boundary conditions are applied. (perfect magnetic conductor in the direction of the source magnetic field and perfect electrical conductor on the orthogonal boundaries). An impedance boundary condition at 1 S/m represents the ocean interface at these higher frequencies that do not fully penetrate the ocean. The source field is specified to be horizontal at an altitude of 10 km. A typical element size of the unstructured tetrahedral mesh in the crust is 1 km, which is refined to include at least 5 elements across the targets.

The MT response was evaluated for a dike and a magma chamber, in addition to checking the 3D calculation for the sill against its 1D result. The dike is the sill on end, i.e., an 0.2-km thick sheet displaced 2 km horizontally from the observation point. The magma chamber is represented as a sphere is centered at 2-km range and 1-km depth and is 1.9 km in diameter (so it does not touch the surface). The conductivity of all three crustal targets is 0.1 S/m. The fractures and putative source pit on Vinalia Faculae guided the choice of these geometries (Fig. 4 in Scully et al., 2020).

To better visualize the results, the apparent conductivities of the three crustal brine intrusions are normalized to the apparent conductivity of the ocean alone (**Fig. S2**). The response of the sill is isotropic (identical in TE and TM) and has the largest departure from the background response at ~0.3 Hz (see also Fig. 3). The dike is similarly distinguished in the TE mode but not in TM: the latter treats the part of the electric field that is perpendicular to the dike, which is not strongly perturbed from the ocean response. In either TM or TE, the apparent conductivity of the structure including the sphere is nearly a constant ratio to that of the ocean alone. This is a manifestation of "static shift" common in terrestrial MT, in which shallow, unresolved conductors uniformly displace the responses of deeper structure. Not only is the shallow conductor unresolved, but the deeper structures are erroneously mapped. In this example, however, the anisotropic response of the sphere (<1 in TM, >1 in TE) signals the presence of the shallow heterogeneity and would allow the ocean depth to be accurately recovered. While comprehensive study is necessary to determine the full capabilities of MT in assessing shallow intrusions on Ceres, this preliminary work demonstrates that simple structures can be distinguished from each other and from the ocean by their anisotropic responses.

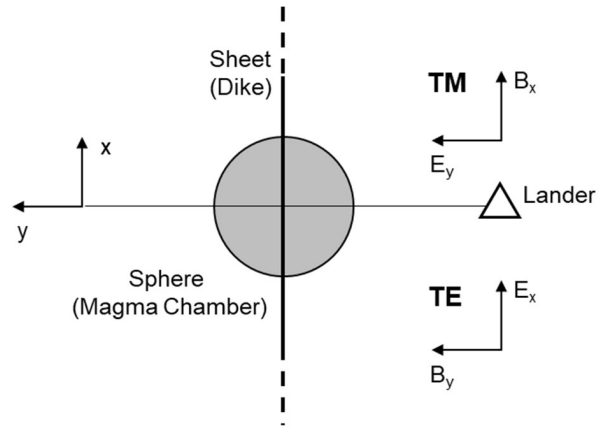

**Fig. S1.** Surface projection of the geometry for 3D calculations, including definition of TM and TE modes.

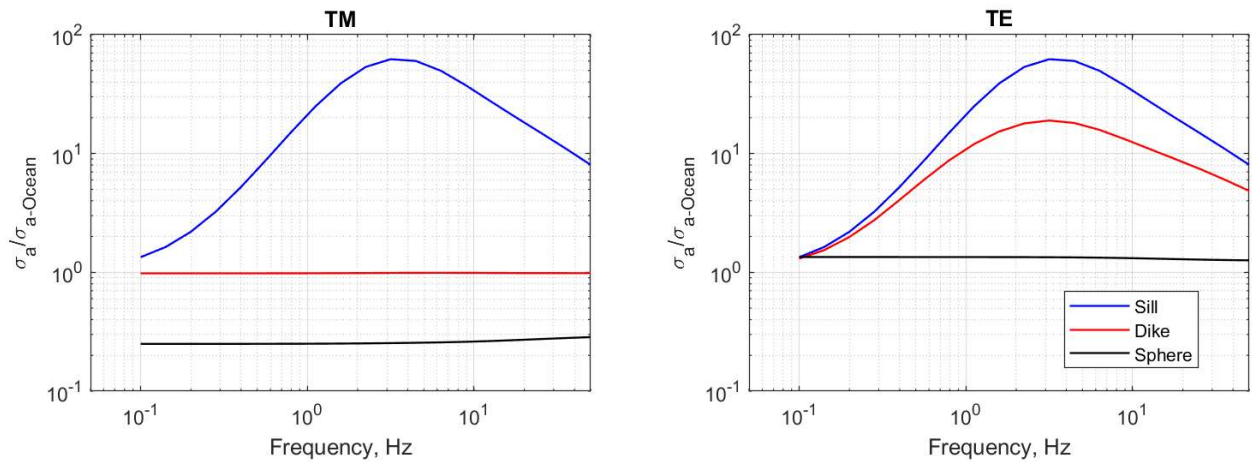

**Fig. S2.** 3D calculations of a vertical dike (offset 2 km, width 0.2 km) and a sphere (offset 2 km, diameter 1.9 km, depth 1 km) compared to reference horizontal sill (depth 2 km, thickness 0.2 km). Intrusion conductivity 0.1 S/m in all cases. Curves are normalized by the response of the 1-S/m ocean at 35-km depth. Anisotropy of TE and TM modes enables different structures to be distinguished.